# Use of UML and Model Transformations for Workflow Process Definitions

Audris Kalnins, Valdis Vitolins[1]
*University of Latvia, IMCS, 29 Raina blvd, LV-1459 Riga, Latvia*

**Abstract.** Currently many different modeling languages are used for workflow definitions in BPM systems. Authors of this paper analyze the two most popular graphical languages, with highest possibility of wide practical usage - UML Activity diagrams (AD) and Business Process Modeling Notation (BPMN). The necessary in practice workflow aspects are briefly discussed, and on this basis a natural AD profile is proposed, which covers all of them. A functionally equivalent BPMN subset is also selected. The semantics of both languages in the context of process execution (namely, mapping to BPEL) is also analyzed in the paper. By analyzing AD and BPMN metamodels, authors conclude that an exact transformation from AD to BPMN is not trivial even for the selected subset, though these languages are considered to be similar. Authors show how this transformation could be defined in the MOLA transformation language.

**Keywords**: workflow, BPM, UML Activity Diagram, BPMN, metamodel, model transformation

## Introduction

Though we know the workflow concept for almost twenty years, the recent emergence of business process management (BPM) systems for enterprises has completely revolutionized the role and place of workflows. Classical workflow systems were intended for pure document processing. In recent years workflow systems have converged with enterprise resource planning (ERP) and Enterprise Application Integration (EAI) systems, and have become completely distributed and global – they are the main driving force for many B2B systems, closely integrating with web services. It has added so many new aspects for workflow design, but at the same time the main feature which distinguishes workflows from other software systems – to support humans in their business activities remains valid.

At this time there is a bunch of organizations (W3C, WfMC, OMG, OASIS, AIM, etc.) and standards (UML [1], BPMN [2], BPEL [3], WSCL, WSCI, ebXML), which standardize some aspects of workflows and provide different design notations.

Only two of these standards - UML Activity Diagrams (AD) and Business Process Modeling Notation (BPMN) provide an easy readable graphical notation for workflow processes [1,2]. Some authors assert that ADs and BPMN are similar [15] and there is a new OMG initiative [4,20] to join UML AD and BPMN under one integrated metamodel, though an exact transformation between them is not a straightforward task. Considering also different versions and implementations of workflow execution languages [3,5,6,7], only model driven development (MDD) approach, which would be able to transfer the required workflow design from one notation to another without the loss of meaning, can save the situation.

[1] Supported partially by ESF (European Social Fund)

The paper proposes such an approach using UML 2.0 as a universal platform independent workflow design language. The use of UML activity diagrams is analyzed and a special workflow profile for them is offered (in section 3). BPMN notation [2] is analyzed in the role of another workflow definition language (section 4). Finally, a mapping between these notations is defined and a transformation to BPMN is described in MOLA transformation language [30] (section 5). However, the approach is not limited to this specific choice of target notation, a reverse transformation or a completely different target notation could be treated in a similar way.

## 1. Required Aspects of Workflow Definition

As it was noted in the introduction, the role of workflow in BPM has significantly increased in recent years, however the basic features of a "classic workflow" are still valid. In this paper we are concerned mainly with process definitions, namely, with some development steps of these definitions in MDD style. It is clear that process definitions should be rich, expressive and formal enough to allow workflow process execution by a workflow engine, as the final step of this development.
Considering workflow system requirements [13,14] and requirements from workflow pattern definitions [16,17,18,19], we can nominate the following "workflow complete" requirements for workflow system definitions:

- Adequate description of the **control structure** of the process – support for branching, decisions, parallelization, synchronization and loops.
- Possibility to model **data flows** and to base decisions, looping and branching on **process data** - "variables associated with a process instance" [12].
- Possibility to define **different types of tasks** - a) for invocation of a workflow subprocesses; b) automated tasks, which is either executed by workflow engine directly (Script Task), or by some sort of automated software service (Service Task, e.g. for Web services), and c) tasks where a human performer performs the task with the assistance of a software application (User Task), or task that is expected to be performed without aid of the workflow system (Manual Task, the role of workflow engine should be in notifying the performer and allowing to signal the task completion.).
- **Resource management** – mainly, the definition and management of human task performers.

## 2. Languages for Workflow Design and their Role

Obviously, workflow definition requires an easy readable language with a clear execution semantics. Most authors agree [10,11,12,13] and it is also our opinion that it must be a graphical language relying on already established business process modeling notations. Currently only two popular candidates are available – UML activity diagrams [1] and BPMN notation initially supported by BPMI [2]. Neither of them had got an adequate industry support yet.

UML 2.0 Activity diagram is the most promising notation for defining workflow processes, due to its precisely defined semantics and the general popularity of UML. Usability of UML 2.0 AD for workflow definition has been confirmed by several authors [14,15,17,18,23,24,25], several deficiencies also being noted at the same time. This analysis is mainly based on the pattern approach [15,16,17,18,19] (introduced by W. van der Aalst et al.), initially the control flow patterns, but afterwards also data and resource patterns. According to this approach, a list of patterns (inspired mainly by practical cases) is selected and it is checked, which of the patterns can be easily implemented in the given notation – here AD. It should be mentioned that the pattern lists include also some very complicated patterns not showing up so frequently.

The approach in this paper is in a sense complementary – we want to provide a pragmatic complex solution how AD could be used to cover the above mentioned aspects of workflow definition in typical cases, not separating so strictly control and data aspects (this way, e.g., the *Synchronising merge* pattern [17,18] not supported formally by AD can be implemented in any practical case).

Another reason, why we see UML 2.0 as the one of the most viable candidate for process definition is that its class diagram supports a precise definition of enterprise domain model and various software interfaces, including web services. As a consequence, UML activity diagrams for the process definition can rely on rich data structures defined by the class model and this is illustrated in our approach.

Our goal is to offer a practical AD profile for workflow definition which is executable through mapping to some executable language. In other words, we try to create our own DSL (domain specific language) as a profile of the UML – the introduced stereotypes should hide the found AD deficiencies for practice.

Namely for workflow definition another well known graphical language - BPMN [2,19] has been proposed. It has got also certain support in tools [9]. But this language has its own set of deficiencies, especially the informal semantics and lack of adequate features for data definition. Though a mapping of BPMN to BPEL is defined in its standard [2], this mapping is quite informal. In section 5 the usability of BPMN for workflow definition is briefly analyzed, in order to find a relevant subset.

Another new idea this paper proposes is that in situation with no one leading notation only the MDD (model transformation) approach can save the situation. In section 5 we show the first MDD step - how a UML AD model can be transformed to a semantically equivalent BPMN model. The executable transformations are defined in the model transformation language MOLA [30].

In a typical MDD approach workflow definition life cycle consists of several transformation steps from more platform independent languages to more specific languages. Situation with platform specific process notations is also quite complicated. There are several vendor-specific process definition languages (IBM WBI [21], Oracle BPEL [7], xBML, XLANG). However, the most popular and supported by industry is the BPEL [3] language - currently the version 1.1 [3], with version 2.0 [5] coming. It is implemented by several workflow engines provided by leading workflow vendors [6,7]. The BPEL language standard covers only the software integration aspect according to SOA (invoked web services). However, BPEL engines by leading vendors – IBM and Oracle [6,7] provide also some aspects of human resource control and manual task support, as vendor specific extensions. Thus, some version of BPEL is the most probable PSM for workflow development. Several tools [9] already propose transformations from BPMN to BPEL, though not flexible enough.

Some papers [23,24,25,26,28] propose direct transformations from UML AD to a BPEL dialect, but none of them uses a UML subset/profile rich enough to cover the workflow aspects discussed in this paper. No doubt that a direct transformation from AD to BPEL could be defined in MOLA in a flexible way. But this is a topic of another paper.

## 3. Adjusting UML Activity Diagrams for Workflow Definition

UML activity diagrams (AD) serve as the main graphical process design notation (a sort of PIM) according to our approach. In this section we specify the recommended subset of AD notation and propose an AD profile for process definition. Several AD profiles for defining workflows have already been proposed [23,25], but none covers all the required features. The recent IBM WBI [21] process notation (not offered as an AD profile) actually is similar to the UML 2.0 AD subset proposed by us.

The role as a PIM for a complete precise workflow definition puts new requirements on AD. The main issues are: an adequate choice of action subtypes in AD, use of data for the precise execution logic definition and the definition of action performers.

### *3.1. AD subset for workflow definitions*

AD notation has a vast majority of features most of which make no sense for workflow definition, because they have been introduced for completely different purposes (defining executable processes for embedded systems). Therefore we select a list of recommended AD features for workflow definition, retaining the original semantics of AD elements. Stereotypes introduced later (section 3.2) will add some missing properties and constraints required for workflow practice (or clarify semantics in places where the original semantics is not precise enough). As a rationale for choice the executable semantics of workflow elements, which most clearly appears in BPEL, was used.

Fig. 1 shows a typical B2B workflow example containing two activities. Each activity (which corresponds to an AD in the graphical form) represents a workflow process which is executed in one organization (i.e., on its own workflow engine). `Customer Process` is started externally, but `Supplier Process` is started when a new `Order` is received. In others words, the whole `Supplier Process` acts as a new web service.

We start with **actions**, which should cover all task types mentioned in the section 1. Because action equivalent in the workflow domain is **task,** both these terms will be used as synonyms. Only few action types have sense for workflow definition. The **call behavior action** will be used for denoting subprocess invocations, manual, user and script tasks. **Call operation action** will be used for denoting service tasks, since there an operation implemented by a software component (web service) is invoked. In the web services world **send signal** action actually is a kind of asynchronous **call operation** (it has the same implementation), but **accept event** (including accept timer) action retains its AD role and usage.

UML AD has sufficient facilities for defining guard conditions (e.g., in OCL) for decisions and other edges, the question is where the referenced data come from. The use of object flows everywhere does not correspond to usual workflow thinking, where each workflow process normally has variables. Therefore we recommend the use of **variables** in AD, together with **read variable** and **write variable** actions manipulating them. A variable can be scoped by the activity itself and by a structured activity node (from which we use only one stereotyped subkind – loop). Neither variable definitions nor read/write variable actions have a graphical notation in the standard. We propose explicit action boxes with assignment operation in the action body (in the form y:=f(x)).

Control flows, object flows and all kinds of control nodes will be used. For representing object flows we recommend the use of pin notation, and variables for proper data transformations.

According to UML semantics, if **joins** combine different types of object flows, then all these objects (tokens) are passed to next action. Since in workflows frequently only one of the joined objects is used by the next action (the other is used only in the join condition), we specify this by the type of the receiving pin.

We recommend **interruptible** regions with interrupting edges for exceptions, cancellations etc. **Loop** nodes will be used with the test section of a special form (`ForEach <element> IN <set>`) and arbitrary graphical body section. **Partitions** will be used for defining action performers via `represents` association.

Due to workflows becoming an integral part of complicated business processes, there is a necessity to define workflows between different organizations. We describe such orchestration of several collaborative workflows using several ADs (one per participant), which are transformed to one BPMN diagram containing several processes (pools).

Fig. 1 shows an example of a simple collaborating workflow process definition which is described by two activity diagrams – the `Customer Process` and `Supplier Process` AD.

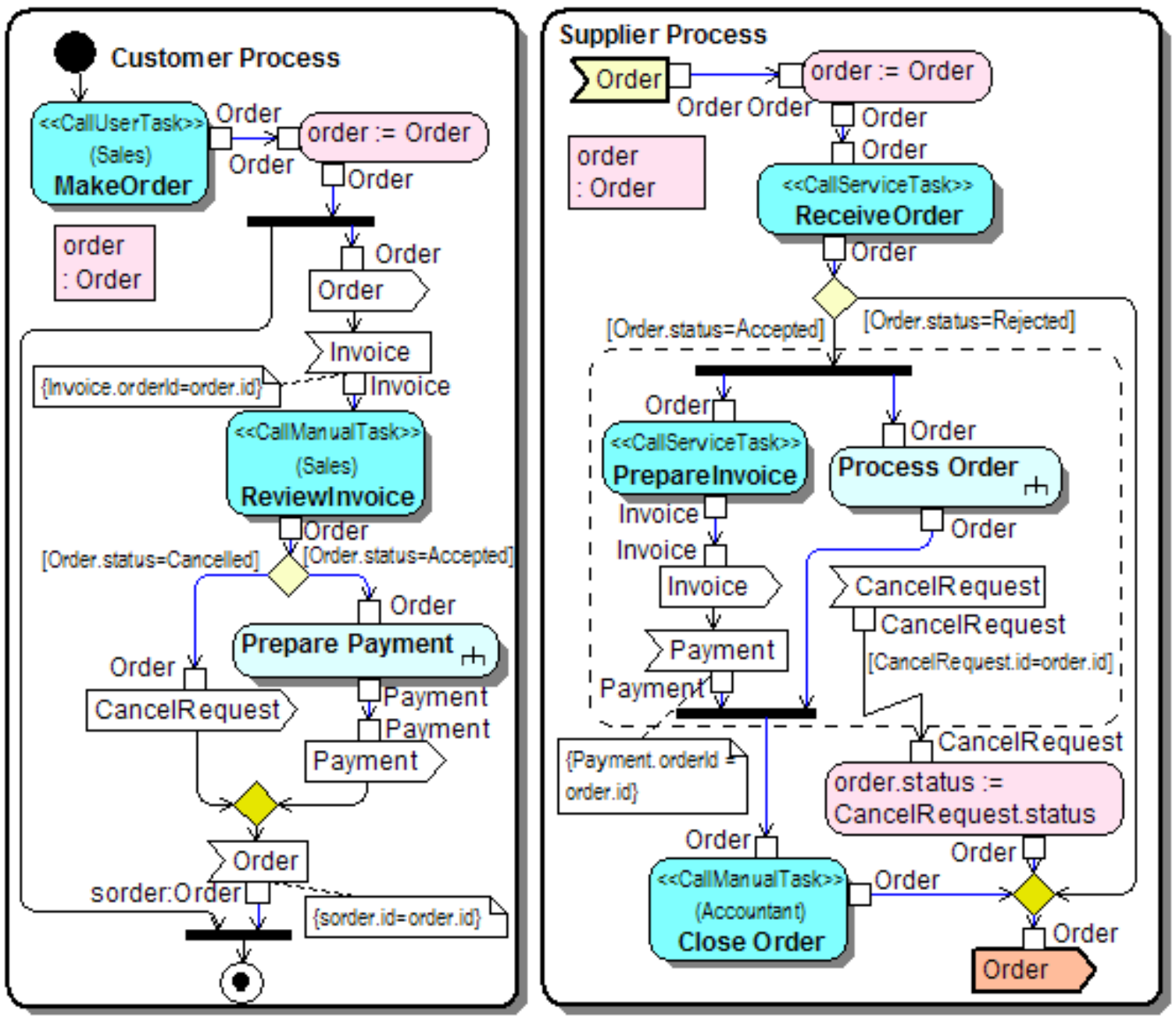

**Fig. 1.** MainProcess Activity Diagrams with InterruptibleActivityRegion and ActivityPartitions

The `Process Order` subprocess of this process is shown in Fig.2. Nearly all of the recommended AD elements are visible in these examples. Some of the elements have a non-standard graphical representation. The element `order : Order` is a variable definition of type `Order`, with the scope of activity `Supplier Process`. Write variable actions are presented as normal assignments to the variable, simple OCL syntax is used for expressions. Certainly, we rely here on data structures defined in the class model (for the example it is so simple that it can be guessed uniquely).

AcceptEventActions with data based guard conditions for the outgoing object flow is a specific pattern for locating the relevant process instance at message reception, which is equivalent to the use of explicit correlation sets in BPEL [3]. In future the general acceptation of WS Addressing standard will permit to use also an implicit message addressing directly to the relevant process instance [10].

Fig. 2 shows another activity which is a subprocess of Supplier Process. It starts and ends with activity parameters and contains ForEach node representing a loop.

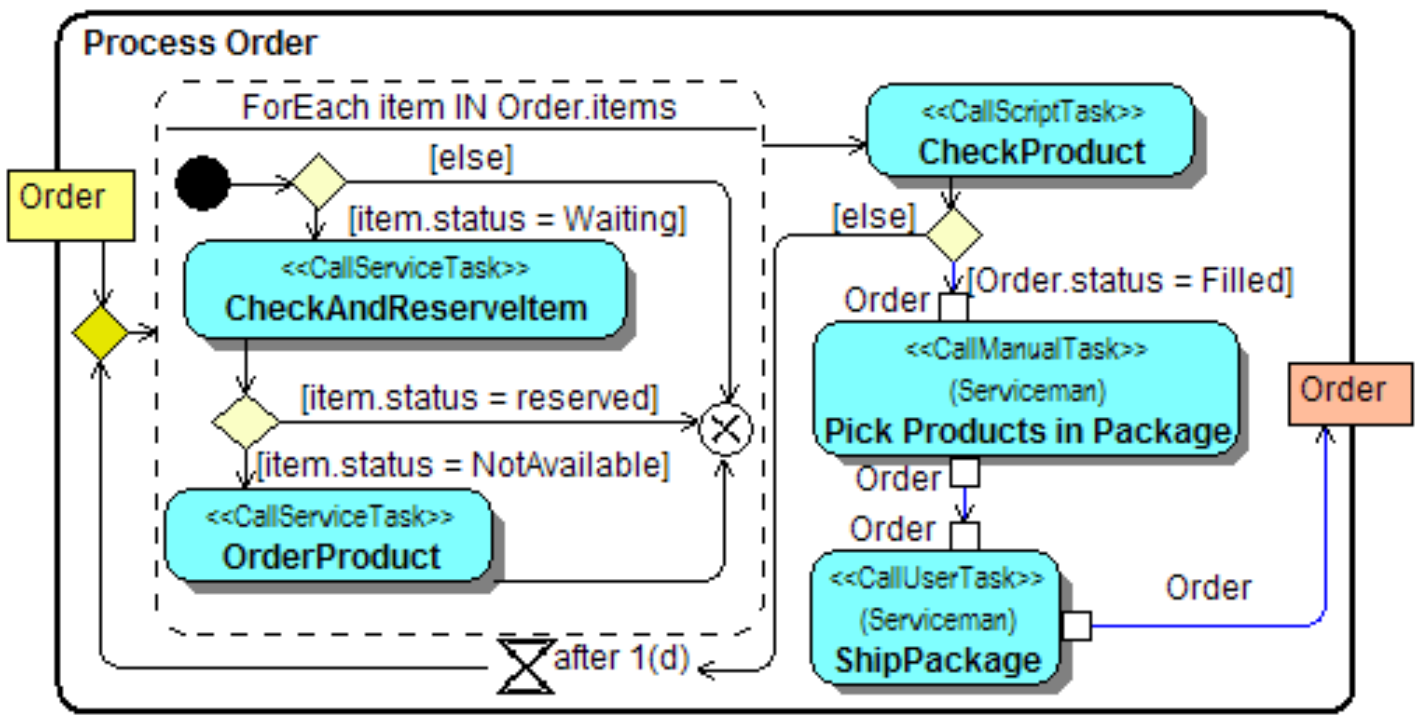

**Fig. 2.** SubProcess Activity Diagram with ForEachNode and ParameterNodes

All AD diagrams are built using the metamodel based GMF (Generic Modeling Framework tool [32], similar to the Vanderbuilt GME [31]), which allows to present element stereotypes graphically in a very flexible way.

The examples illustrate also our proposal (in the next section) for stereotypes of AD elements. Stereotypes for actions are presented in the traditional guillemet style, but other stereotypes are presented by the graphical style of the corresponding symbol.

### *3.2. Workflow profile for AD*

As it was stated in section 3.1, in addition to selection of an appropriate subset, we need a **Workflow Profile**, which describes more adequately the intended workflow-related semantics of chosen AD elements. However, profiles in UML 2.0 is not a powerful enough extension mechanism (a "heavyweight" extension of metamodel is used for the *ForEach* stereotype). The proposed stereotypes could be defined formally (as OCL constraints), but we provide only some informal descriptions due to size limitations (stereotype names are in italic, class names are normal).

***MainProcess*** represents an Activity which is a separate workflow process (executed by an individual workflow engine). Graphically it is shown as a shadowed activity. All technical attributes (e.g., for the relevant BPEL definition header) can be specified as attributes of its corresponding class (each Activity is a Class).

***Performer*** (stereotype for Partition) represents a performer of a manual or user action. Its `represents` association has to reference a class with the *Position* or *OrgUnit* stereotype. As it is noted in [17,18], in such way only a user or role-based allocation is possible, but it is sufficient for most of cases. A *performer* partition is shown as a compartment of the action.

***WebService*** (stereotype for Component) is used to describe web service attributes. *CallServiceTask* actions, which invoke operations within this service, get their technical parameters from here.

***IntermSSAction*** means sending of the specified signal to a web service (it is shown as a simple convex pentagon). ***EndSSAction*** (shown as a convex pentagon with bold border) means sending a signal as the final action of the activity.

The existing UML LoopNode has restricted usability, because its setup, test and body parts can contain only ActivityNodes. Thus it is difficult to define a traditional iterator over a collection. Therefore we introduced ***ForEach*** stereotype for LoopeNode (Fig. 2) with one new `ForEach.collection` association which references the `ValuePin` (Fig. 3).

All other stereotypes are self-descriptive and are shown only in metamodel (Fig. 3), due to size limitation.

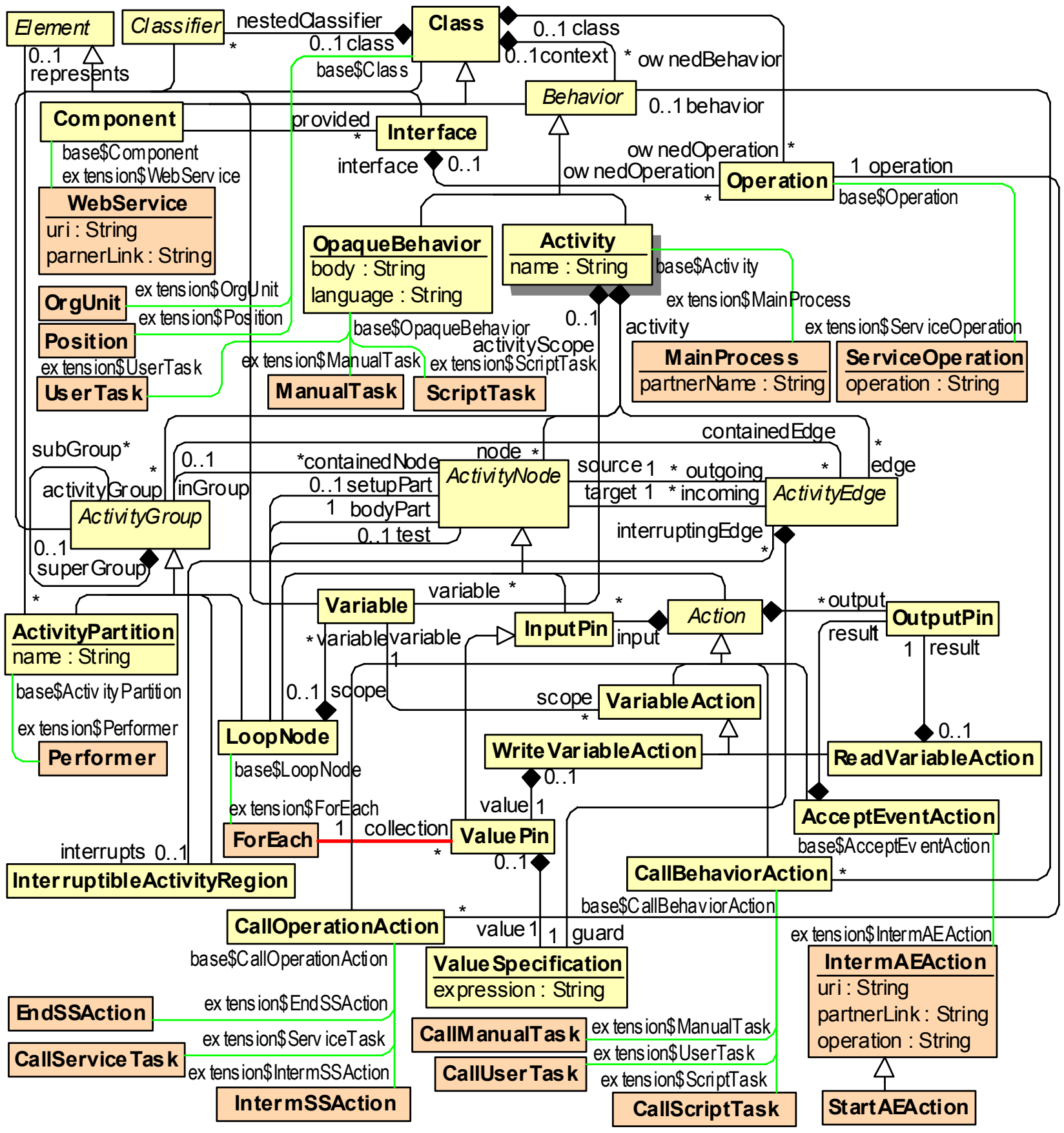

**Fig. 3.** Fragment of the AD Metamodel (source of model transformation)

Fig. 3 shows an "interesting" fragment of "flattened" UML AD metamodel (all redundant abstract superclasses eliminated). According to [1], when our Workflow Profile has been applied to the model, "temporary" metaclasses appear with names equal to stereotype names and also "temporary" associations (e.g., `extension$Position` <-> `base$Class`) to their base metaclasses. Namely this fragment is sufficient to understand the transformation example in section 5.

## 4. BPMN diagrams as another notation

As it was pointed out in introduction, BPMN is also a widely used language for workflow definition. For analysis of BPMN, we rely on the document version 1.0 [2]. No formal BPMN metamodel is available, however it can be "reengineered" in a quite unambiguous way from the textual description. At first, we want to describe briefly the **BPMN subset** used in our paper.

Since our goal is to define **executable** workflow processes, we use only these BPMN elements, which have a mapping to BPEL language [2,27,28]. All kinds of Gateways are used, and also all types of Tasks and Subprocesses. We use Start and End Events, and only IntermediateEvents, attached to the boundary of an Activity ("interrupt construct"), because they all have a natural semantics and mapping to BPEL.

Fig. 4 shows a BPMN process example in the chosen notation – it is a precise equivalent of the process in Fig. 1 (assumed to be the result of the transformation).

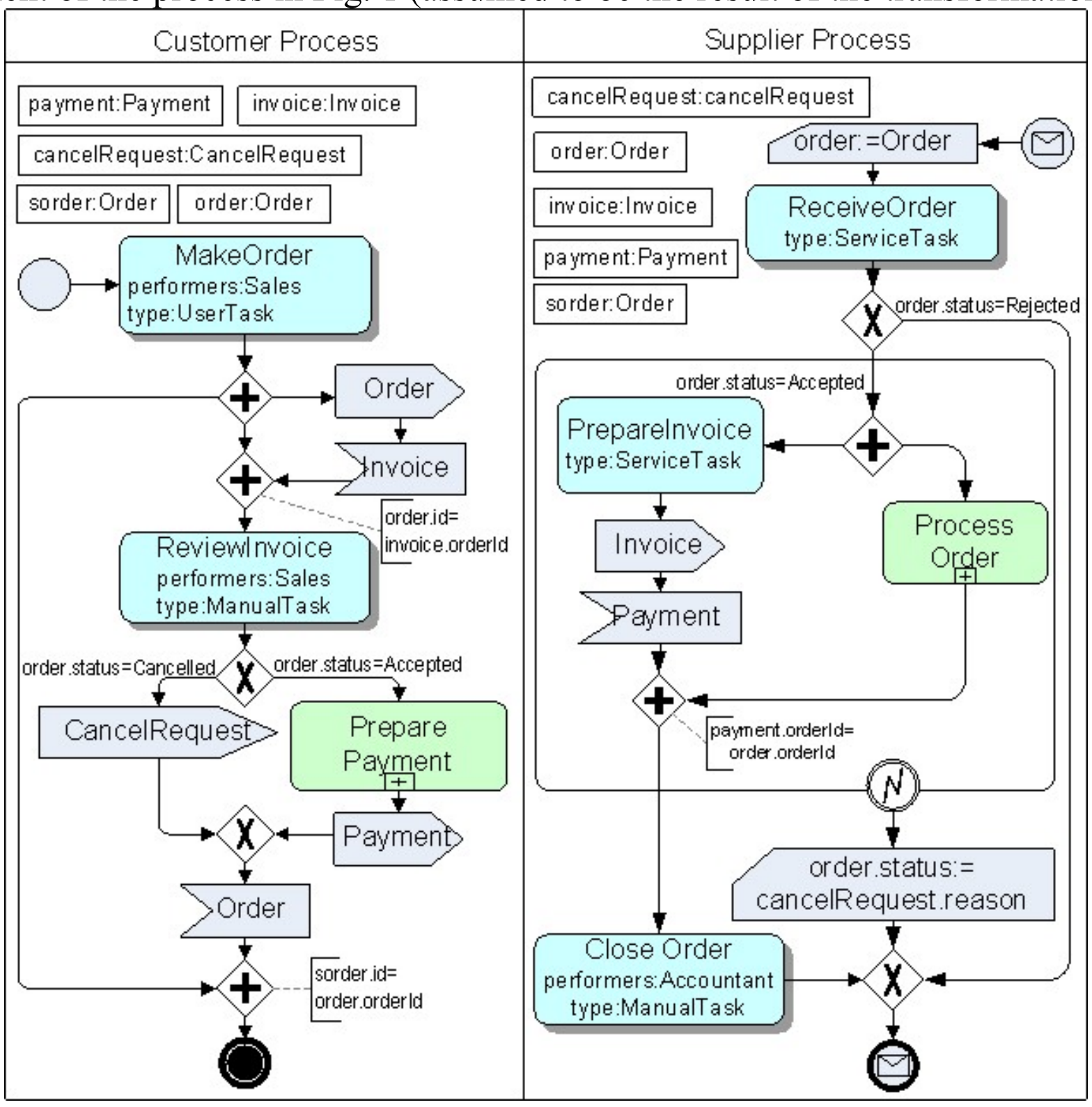

**Fig. 4.** Processes shown as pools, embedded subprocess and exception flow in BPMN

In the original standard [2,27] only Service Tasks are treated as executable, but we treat all types of tasks as executable (basing on an "extended BPEL" similar to [7]), therefore we have extended the BPMN metamodel by moving these associations from ServiceTask to Task.

We don't use Data Objects, MessageFlows and all other Intermediate Events, because they have no unambiguous mapping to BPEL [2,27]. For receiving messages in non-interrupt situations we prefer Receive tasks, and Send tasks for sending accordingly.

While the implicit BPMN metamodel is quite acceptable, the BPMN graphical notation lacks some important elements. Therefore we have introduced some "ad-hoc notation" for these elements. We use explicit tagged-value notation for task type and for task performers. We present Property explicitly in the diagram as a rectangle containing `name:type`. To make the data aspect visible, we show an assignment as a rectangle with upper left corner cut, containing the textual assignment statement. For SendTask and ReceiveTask we use shapes similar to UML – a convex pentagon for Send Task, and a concave one for Receive. Similarly to the AD, data join criteria are shown as TextAnnotations (analog to Notes in AD).

Following the BPMN style, we have not shown input/output Messages for tasks (analogs to AD pins), but we emphasize that we use them as invisible model elements. To maintain diagram readability, the implementation related aspects (e.g., service operation, URI) are not shown, similarly to the AD.

Similarly to UML AD, Fig. 5 shows a "practical" subset "reengineered" from BPMN standard. These elements should appear as the transformation result.

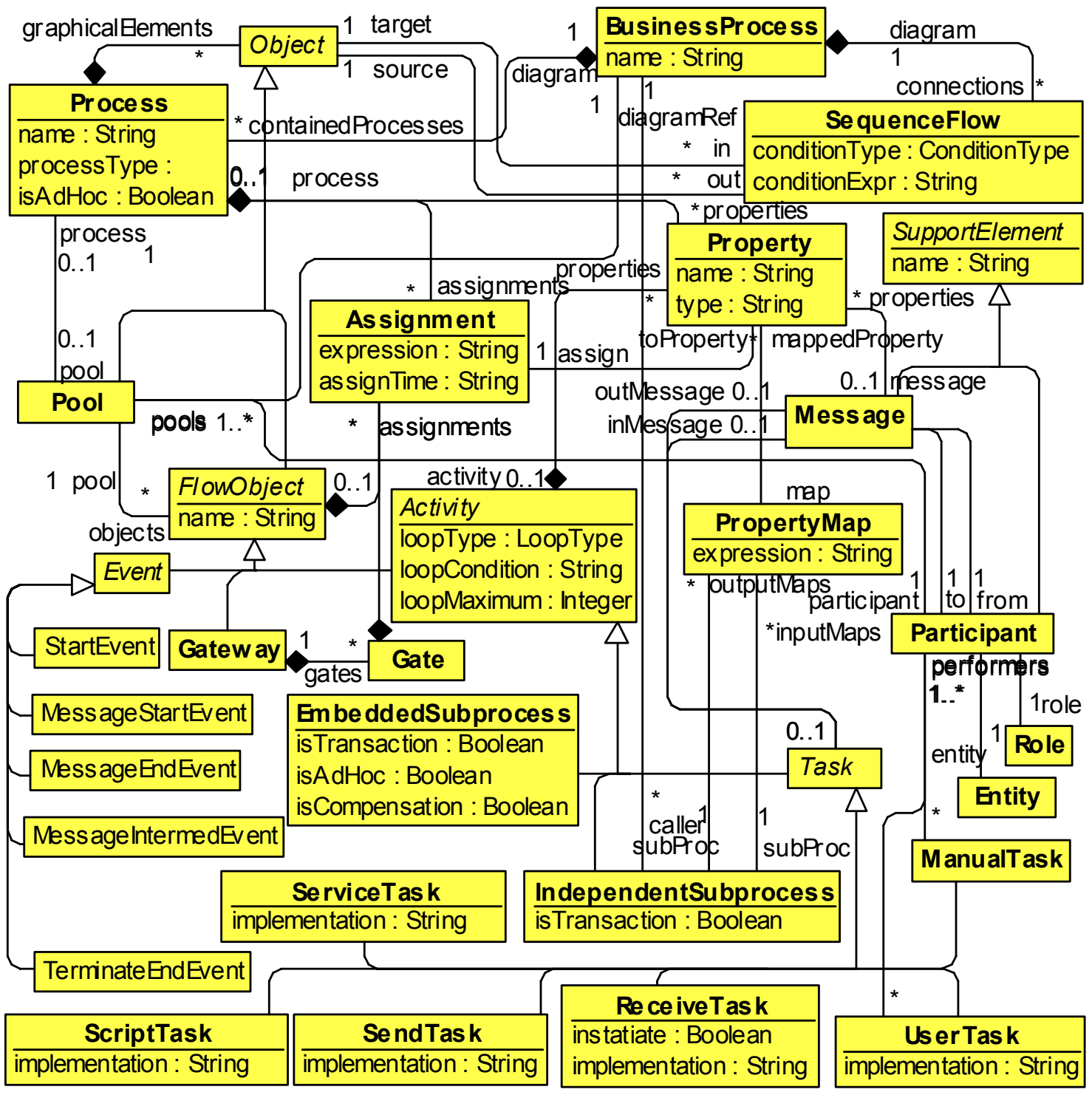

**Fig. 5.** Fragment of the BPMN Metamodel (target for transformation)

## 5. AD to BPMN transformation

There are several works related to the formal process transformations. [8,11] transform UML 1.4 activity graphs to BPEL 1.0. Several papers analyze the possibility to transform directly UML 2.0 AD to BPEL 1.0 [23,24,25,26]. A transformation between BPMN and BPEL (in the form of formalized relations) is described in [27,28]. For few tools [9] hardcoded BPMN to BPEL transformation is already implemented, but it is quite superficial and details are vendor specific.

We have found no existing formalized transformations from UML AD to BPMN supporting all workflow related aspects.

Our goal is to describe a formal **transformation from UML to BPMN** which would support our workflow definition objectives. Namely, a set (model) containing related UML ADs (with supporting class definitions) must be transformed to a similar set of BPMN diagrams, with the same accuracy for control- and data flows, and human performers. The intended semantics (mapping to BPEL) has also to be preserved.

A mapping from AD to BPMN (mainly at the conceptual level, though with exhaustive analysis for control patterns) is done in [19,22]. This mapping of control structures satisfies our goals too, therefore we extend mapping only for data flows, variables, assignments, and task performers. Due to paper limitations the mapping is only briefly shown (incompletely!) as respective AD metamodel classes (or *stereotypes*) and BPMN metamodel classes:

Activity = Process + Pool, *Performer* = Participant, *CallManualTask* + *ManualTask* = ManualTask, *CallServiceTask* + Operation + Interface + *WebService* = ServiceTask, InputPin/OutputPin = Message, ObjectFlow = SequenceFlow , *OrgUnit* = Entity, *Position* = Role, Variable = Property, ReadVariableAction/WriteVariableAction = Assignment, InterruptibleActivityRegion+ ActivityEdge + *IntermAEAction* = EmbeddedSubprocess+ MessageIntermedEvent, … .

To refine the mapping, in this paper we provide an illustrative fragment of the formal transformation generating a BPMN model from an AD model. We note once more that transformations in both directions are of high value, but transformation from AD to BPMN is chosen because it may be more interesting in practice.

The transformation is written in the MOLA language. Due to paper limitations we cannot describe MOLA language syntax therefore for complete reference see [30]. Fig. 3 and Fig. 5 represent the source and target metamodels of the transformation, respectively, but Fig. 6 illustrates the main program of the transformation, which transforms each Activity to a new BPMN Process and Pool, and invokes subprograms transforming all other elements for this Activity.

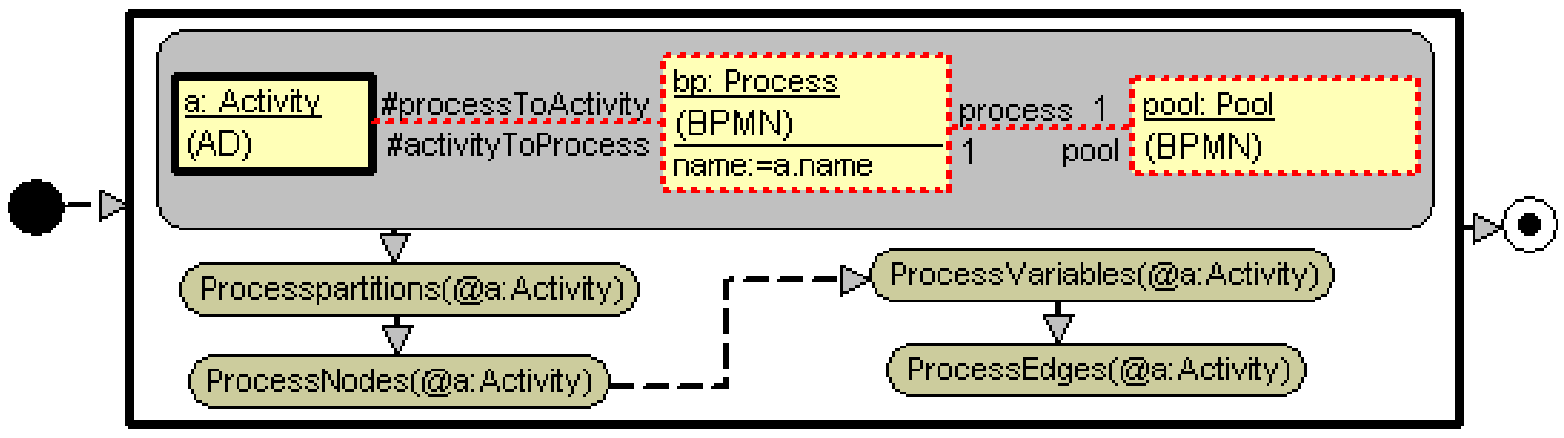

**Fig. 6.** Main transformation program of AD to BPMN

Fig. 7 shows one of the subprograms transforming each AcceptEventAction with its stereotype *IntermAEAction* to the ReceiveTask in the BPMN.

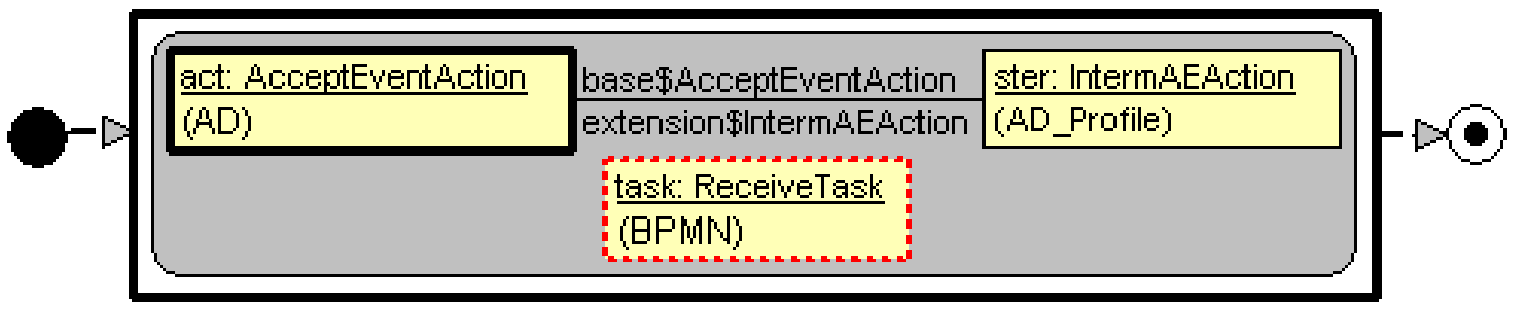

**Fig. 7.** Transformation of AD AcceptEventAction to BPMN ReceiveTask

Namely the MOLA program describes completely formally the mapping between these two notations, which in some cases is quite complicated (involving several classes and context conditions) and cannot be specified by pure descriptions.

The model transformation described here could be defined in any of the model transformation languages, including the MOF QVT language [29]. However, the use of the graphical MOLA language makes the transformations more readable and allow to serve them as formal comments for transformations. In our view, model transformations is the best way how to define a complicated mapping between two notations, including the issues of semantics preservation.

## Conclusion

The paper has demonstrated a possible use of MDD approach in the workflow definition area. Semantics of UML and BPMN languages was analyzed with emphasis on workflow relevant data and manual tasks. Usage of UML profile allowed to offer UML AD as a viable notation for workflow definition, with improved support for integrating control and data aspects and resource (performer) management. The specific subset of BPMN notation was selected in the role of target language. Mappings to BPEL were used as selection criteria for workflow constructs and as a base for semantics refinement. Mapping between AD and BPMN notations and a transformation example in MOLA language (as a refinement of this mapping) were given.

An important aspect is also the tool support. The transformations were built using the MOLA tool [30] developed at the University of Latvia. For fully implementing the complicated UML profile (with graphical extensions), the generic metamodel based modeling tool GMF [32] was used, which is closely integrated with the MOLA tool. Alternatively, activity diagrams (without graphical extensions) could be defined in Rational Software Architect (RSA), where MOLA can be used as plugin. Using the described MDD approach the defined workflows can be transformed to any vendor specific execution language. The exchange with the proper workflow world is possible via the standard XMI coding of models.

## References

[1] Unified Modeling Language: Superstructure, version 2.0, 2005, Object Management Group (OMG), http://www.omg.org/cgi-bin/doc?formal/05-07-04

[2] Business Process Modeling Notation (BPMN), Version 1.0 - May 3, 2004, Business Process Management Initiative (BPMI), http://www.bpmi.org/

[3] Business Process Execution Language for Web Services version 1.1, ftp://www6.software.ibm.com/software/developer/library/ws-bpel.pdf

[4] Andrew Watson, OMG's new modeling specifications, ECMDA-FA 2005, Nuremberg, Germany, November 7-10, 2005, keynote speech

[5] Web Services Business Process Execution Language Version 2.0, Working Draft 01, December 2004, OASIS Open, Inc., http://www.oasis-open.org/apps/org/workgroup/wsbpel/
[6] BPELJ: BPEL for Java technology, IBM Developerworks, http://www-128.ibm.com/developerworks/library/specification/ws-bpelj/
[7] Oracle BPEL Process Manager, http://www.oracle.com/technology/products/ias/bpel/index.html
[8] Draft UML 1.4 Profile for Automated Business Processes with a mapping to BPEL 1.0 http://www-128.ibm.com/developerworks/rational/library/4593.html
[9] BPMI, Current Implementations of BPMN, http://www.bpmn.org/BPMN_Supporters.htm#current
[10] Mike Havey, Essential Business Process Modeling, O'Reily, 2005, ISBN: 0-596-00843-0
[11] Ken Beck, Joshy Joseph, Germán Goldszmidt, Learn business process modeling basics for the analyst, IBM Developerworks, http://www-128.ibm.com/developerworks/library/ws-bpm4analyst/
[12] Tom Baeyens, The State of Workflow, May 2004, http://www.theserverside.com/articles/content/Workflow/article.html
[13] Derek Miers, Paul Harmon, The 2005 BPM Suites Report, Version 1.0, March 15, 2005, http://www.bptrends.com/reports_toc_01.cfm
[14] Curtis Hall, Paul Harmon, The 2005 Enterprise Architecture, Process Modeling & Simulation Tools Report, Version 1.0, April 28, 2005, http://www.bptrends.com/reports_toc_02.cfm
[15] Stephen A. White, Process Modeling Notations and Workflow Patterns, BPTrends March, 2004, http://www.omg.org/bp-corner/pmn.htm
[16] N. Russell, W.M.P.van der Aalst, A.H.M. ter Hofstede, and D. Edmond. Workflow Resource Patterns: Identification, Representation and Tool Support., Proc. of the 17th Conference on Advanced Information Systems Engineering, vol. 3520 of LNCS, 216-232. Springer, Berlin, 2005.
[17] P. Wohed, W. M.P. van der Aalst, M. Dumas, A. H.M. ter Hofstede, N. Russell, Pattern-based Analysis of UML Activity Diagrams, BETA Working Paper Series, WP 129, Eindhoven University of Technology, Eindhoven, 2004
[18] N. Russell, W.M.P. van der Aalst, A.H.M. ter Hofstede, and P. Wohed.. On the Suitability of UML 2.0 Activity Diagrams for Business Process Modelling. BPM Center Report BPM-06-03, BPMcenter.org, 2006.
[19] P. Wohed, W.M.P. van der Aalst, M. Dumas, A.H.M. ter Hofstede, and N. Russell. Pattern-based Analysis of BPMN - An extensive evaluation of the Control-flow, the Data and the Resource Perspectives. BPM Center Report BPM-05-26, BPMcenter.org, 2005.
[20] Business Modeling & Integration Domain Task Force, http://bmi.omg.org/
[21] WebSphere Business Integration Modeler, IBM, http://www-306.ibm.com/software/integration/wbimodeler/
[22] Business Process Definition Metamodel, Revised Submission to BEI RFP bei/2003-01-06, Object Management Group (OMG), http://www.omg.org/docs/bei/04-01-02.pdf
[23] Behzad Bordbar, Athanasios Staikopoulos, On behavioural model transformation in Web services, 5th International Workshop on Conceptual Modeling Approaches for e-Business eCOMO'2004, November 8-12, 2004
[24] Jean Bézivin, Slimane Hammoudi, Denivaldo Lopes,Frédéric Jouault, Applying MDA Approach to B2B Applications: A Road Map, Workshop on Model Driven Development (WMDD 2004) at ECOOP 2004, Springer-Verlag, LNCS, vol. 3344, June 2004
[25] Roy Grønmo, Michael C. Jaeger, Model-Driven Semantic Web Service Composition, 12th Asia-Pacific Software Engineering Conference (APSEC), Taipei, Taiwan. December 2005
[26] Tracy Gardner, UML Modelling of Automated Business Processes with a Mapping to BPEL4WS, Apr 21, 2004, http://www-128.ibm.com/developerworks/rational/library/4593.html
[27] Stephen A. White, Using BPMN to Model a BPEL Process, BPTrends, 3(3), March 2005, 1-18
[28] Chun Ouyang, Marlon Dumas, Stephan Breutel, and Arthur H.M. ter Hofstede, BPM, 2005, Translating Standard Process Models to BPEL, BPM Center Report BPM-05-27, BPMcenter.org, 2005, http://is.tm.tue.nl/staff/wvdaalst/BPMcenter/reports/2005/BPM-05-27.pdf
[29] Meta Object Facility (MOF) 2.0 Query/View/Transformation Specification. OMG Document, ptc/2005-11-01, http://www.omg.org/docs/ptc/05-11-01.pdf
[30] MOLA Model Transformation language, Institute of Mathematics and Computer Science University of Latvia, http://mola.mii.lu.lv/
[31] Ledeczi A. et al., The Generic Modeling Environment, Workshop on Intelligent Signal Processing, Budapest, Hungary, May 17, 2001
[32] L.Lace, E.Celms, A.Kalnins. Diagram definition facilities in a generic modeling tool, 2003,- Proceedings of International Conference on Modelling and Simulation of Business systems, Vilnius, 2003, pp.220-224.